\documentclass[a4paper,11pt]{article}

\pdfoutput=1

\usepackage{a4wide}
\usepackage{doi}
\usepackage{hyperref}

\usepackage{graphicx}%
\usepackage{multirow}%
\usepackage{amsmath,amssymb,amsfonts}%
\usepackage{amsthm}%
\usepackage{mathrsfs}%
\usepackage[title]{appendix}%
\usepackage{xcolor}%
\usepackage{textcomp}%
\usepackage{manyfoot}%
\usepackage{booktabs}%
\usepackage{algorithm}%
\usepackage{algorithmicx}%
\usepackage{algpseudocode}%
\usepackage{listings}%
\usepackage[switch]{lineno} % Use [pagewise] to reset line numbers on each page, or [switch] to toggle line numbers on and off.

\newcommand{\caa}[1]{\textcolor{black}{#1}} % Carlos
\newcommand{\ca}[1]{\textcolor{black}{#1}} % Carlos
\begin{document}

\begin{center}   
\textbf{\LARGE Aerodynamic roughness of rippled beds under active saltation at Earth-to-Mars atmospheric pressures}

\vspace*{0.2cm}

C.A. \textsc{Alvarez}$^a$,
M.G.A. \textsc{Lap\^otre}$^a$,
C. \textsc{Swann}$^b$,
R.C. \textsc{Ewing}$^c$,
P. \textsc{Jia}$^d$
and P. \textsc{Claudin}$^e$
\end{center}

{
\renewcommand{\baselinestretch}{0.9}
\footnotesize
\noindent
$^a$ {Department of Earth \& Planetary Sciences, Stanford University, Stanford, California, USA}\\
$^b$ {RCOAST, New Orleans, Louisiana, USA}\\
$^c$ {Astromaterials \& Exploration Science Division, NASA Johnson Space Center, Houston, Texas, USA}\\
$^d$ {School of Science, Harbin Institute of Technology, Shenzhen, P.R. China}\\
$^e$ {Physique et M\'ecanique des Milieux H\'et\'erog\`enes, CNRS, ESPCI Paris, PSL Research University, Universit\'e Paris Cit\'e, Sorbonne Universit\'e, Paris, France}
\par
}

%%%%%%%%%%%%%%%%%%%%%%
%\linenumbers
%
%\title[Article Title]{Aerodynamic roughness of rippled beds under active saltation at Earth-to-Mars atmospheric pressures}
%
%\author*[1]{\fnm{Carlos A.} \sur{Alvarez}}\email{calvare7@stanford.edu}
%
%\author[1]{\fnm{Mathieu G.A.} \sur{Lapôtre}}
%
%\author[2]{\fnm{Christy} \sur{Swann}}
%%\equalcont{These authors contributed equally to this work.}
%
%\author[3]{\fnm{Ryan C.} \sur{Ewing}}
%
%\author[4]{\fnm{Pan} \sur{Jia}}
%
%\author[5]{\fnm{Philippe} \sur{Claudin}}
%
%
%\affil[1]{\orgdiv{Department of Earth \& Planetary Sciences}, \orgname{Stanford University},\orgaddress{ \city{Stanford}, \state{California}, \country{USA}}}
%
%\affil[2]{\orgdiv{RCOAST}, \orgaddress{ \city{New Orleans}, \state{Louisiana}, \country{USA}}}
%
%\affil[3]{\orgdiv{Astromaterials \& Exploration Science Division}, \orgname{NASA Johnson Space Center},\orgaddress{ \city{Houston}, \state{Texas}, \country{USA}}}
%
%\affil[4]{\orgdiv{School of Science}, \orgname{Harbin Institute of Technology},\orgaddress{ \city{518055 Shenzhen}, \country{P.R. China}}}
%
%\affil[5]{\orgdiv{Physique et Mécanique des Milieux Hétérogènes}, \orgname{UMR 7636 CNRS – ESPCI Paris – Université PSL – Universit\'e Paris Cit\'e – Sorbonne Université},\orgaddress{ \city{Paris}, \country{France}}}
%
%
%\maketitle
%%%%%%%%%%%%%%%%%%%%%%

%\section*{Abstract}
\begin{abstract}
As winds blow over sand, grains are mobilized and reorganized into bedforms such as ripples and dunes. In turn, sand transport and bedforms affect the winds themselves. These complex interactions between winds and sediment render modeling of windswept landscapes challenging. A critical parameter in such models is the aerodynamic roughness length, $z_0$, defined as the height above the bed at which wind velocity predicted from the log law drops to zero. In aeolian environments, $z_0$ can variably be controlled by the laminar viscous sublayer, grain roughness, form drag from bedforms, or the saltation layer. Estimates of $z_0$ are used on Mars, notably, to predict wind speeds, sand fluxes, and global circulation patterns; yet, no robust measurements of $z_0$ have been performed over rippled sand on Mars to date. Here, we measure $z_0$ over equilibrated rippled sand beds with active saltation under atmospheric pressures intermediate between those of Earth and Mars. Extrapolated to Mars, our results suggest that $z_0$ over rippled beds and under active saltation may be dominated by form drag across a plausible range of wind velocities, reaching values up to 1 cm—two orders of magnitude larger than typically assumed for flat beds under similar sediment transport conditions.
\end{abstract}

%~\newline

%\vspace*{0.5cm}
\begin{center}
Nature Communications \textbf{16}, 5113 (2025).\\
\href{https://doi.org/10.1038/s41467-025-60212-7}{\texttt{doi.org/10.1038/s41467-025-60212-7}}
\end{center}
%\vspace*{0.5cm}

\section*{Introduction}

Interactions between winds and sediment occur over a wide range of scales, from the mobilization of individual sand grains to the formation of bedforms and overall landscape evolution. Such interactions are governed by the shear stress imparted by winds onto the sediment bed; in turn, the bed's response to shearing winds affects the wind profile, and thus, bed stresses \cite{kok2012}. Determining basal wind stresses is thus critical to predicting sand fluxes and landscape evolution in response to winds across planetary bodies.

The shear stress imparted by winds onto a surface, $\tau_b$, is controlled by a length scale, known as the aerodynamic roughness length, $z_0$. This length scale is the height at which wind speed is predicted to become null under neutral atmospheric conditions \ca{-that is, when temperature-driven buoyancy effects are negligible-}such that

\begin{equation}
    u(z) = \frac{u_*}{\kappa} \, ln \left(\frac{z}{z_0} \right),
    \label{eq:log_law}
\end{equation}

\noindent where $u(z)$ is the vertical wind velocity, $u_* = \sqrt{\tau_b/\rho}$ is the shear velocity (with $\rho$ the atmospheric density), and $\kappa \approx$ 0.41 is the von Kármán constant. The parameter $z_0$ can also be interpreted as a mixing length governing turbulent fluctuations over a solid surface \cite{jia2023}. It plays an essential role in boundary layer dynamics and controls energy exchanges between planetary surfaces and overlying atmospheres \cite{stull1988}. For example, characterizations of $z_0$ are required to understand how topography, vegetation, and other obstacles impact wind patterns on Earth \cite{finnigan1988, raupach1992, kaimal1994, lancaster1998, king2006}. It also plays a crucial role in regulating the vertical exchange of heat between the surface and the atmosphere \cite{crago2012, kent2017}.

The value of $z_0$ varies with the characteristics of planetary surfaces. Over sand beds with a typical length scale, $r$, of bed roughness elements (such as individual grains or bedforms), one can define a roughness-based Reynolds number,

\begin{equation}
    \text{Re}_{r}= \frac{ru_*}{\nu},
    \label{eq:Rek}
\end{equation}

\noindent where $\nu$ is the kinematic viscosity of the atmosphere \cite{courrech2024}. When Re$_r \lesssim 10$, the bed is said to be aerodynamically smooth. In the absence of saltation, $z_0$ for aerodynamically smooth beds is controlled by the thickness of the laminar sublayer \cite{garcia2008}, such that

\begin{equation}
    z_0 = \frac{\nu}{9u_*}.
    \label{eq:sublayer}
\end{equation}

In turn, when Re$_r \gtrsim 100$ and sand is immobile, the bed is said to be aerodynamically rough, and $z_0$ is controlled by $r$. For flat beds (i.e., in the absence of bedforms), $r$ scales with grain size. Ref. \cite{nikuradse1933} determined experimentally that $z_0$ is given by $k_s/30$, where $k_s$ is the equivalent sand-grain roughness height. For mixed grain populations, $k_s \approx n d_{i}$, where $n$ is of the order 2-3, and $d_{i}$ is the \textit{i}${\text{th}}$ percentile of the grain size distribution (often $d_{50}$, $d_{65}$, or $d_{95}$) \cite{yalin1972, einstein1950, meyer1948}. Thus,  

\begin{equation}
    z_0 \approx \frac{d_{50}}{12}.
    \label{eq:z0_rough}
\end{equation}

As grains are lifted from the surface by winds and saltate (following a series of short, ballistic trajectories before impacting the surface again), a saltation layer forms, which reduces the wind speed near the surface, steepening the velocity profile \cite{bagnold1941, owen1964}. This feedback between saltation and the wind profile prevents the saltation layer from growing unchecked at an exponential rate to any height above the surface \cite{farrel2006}. In this case, $z_0$ is influenced by the thickness of the saltation layer. 

Under terrestrial conditions, the thickness of the saltation layer was shown to relate to the so-called focal point, defined as the vertical height above which saltation does not affect the wind profile \cite{bagnold1941}. Below the focal height, $H_{\text{f}}$, the average upward grain velocity is independent of wind shear velocity, $u_*$. At the focal point, the wind velocity, $U_{\text{f}}$, scales with the threshold shear velocity, $u_{*t}$. Considering the continuity of the velocity profile at the focal point, the aerodynamic roughness length imparted by active saltation can be expressed as

\begin{equation}
    z_0 = H_{\text{f}} \, \exp{\left(- \kappa \frac{U_{\text{f}}}{u_*}\right)}.
\label{eq:z0_Hf}
\end{equation}

However, there is no consensus on how parameters such as grain size affect $H_{\text{f}}$ and $U_{\text{f}}$. For example, it is unclear whether $H_{\text{f}}$ scales linearly with $d$ \cite{duran2019, pahtz2021}, with $\sqrt{d}$ \cite{andreotti2004,duran2006}\ca{, or not at all \cite{pahtz2021}. If one of the latter two models is correct}, then dimensional analysis dictates that another parameter, likely the kinematic viscosity of air, $\nu$, also controls the position (and possibly the very existence) of the focal point \cite{andreotti2004, duran2006, pahtz2021}; Methods). In the following, we test \ca{two different formulations for $H_\text{f}$ \cite{andreotti2004, pahtz2021}, which both account}, at least partially, for variations in atmospheric density \ca{through kinematic viscosity} (Methods).

Whereas no accepted analytical expression exists to date for the focal height and velocity, it was shown experimentally that in the presence of active saltation, the aerodynamic roughness follows an empirical relationship of the form

\begin{equation}
    z_0 = \frac{C \, u_*^2}{2 \, g},
    \label{eq:z0_mod}
\end{equation}

\noindent where $C$ is a constant that varies with the type of surface or environment \cite{charnock1955, owen1964, sherman1992}. In this study, we assume $C$ = 0.01 \cite{sherman2008}.

Through saltation, the bed often self-organizes into bedforms, such as ripples or dunes. These bedforms disturb the flow, increasing form drag. In aeolian systems featuring multiple bedform scales, both skin friction from grain roughness and form drag from all scales of bedforms affect $z_0$. To estimate $z_0$ in such complex conditions, Ref. \cite{jia2023} developed a semi-analytical model (here referred to as Jia et al., 2023; Methods) for multiscale roughness, which includes skin friction and form drag from two scales of bedforms through looped calculations (Methods). However, this model does not capture the effect of active saltation over a rippled bed.

On Earth, wind flows are typically aerodynamically rough, and $z_0$ is controlled by either the surface roughness (below the transport threshold, $u_{*t}$) or the thickness of the transport layer (above it) \cite{field2018}. In turn, surface roughness compounds both skin friction (Eq. \ref{eq:z0_rough}) and form drag (\cite{jia2023}). Whereas models for $z_0$ are relatively well established for Earth (Eqs. \ref{eq:sublayer}-\ref{eq:z0_mod}), several factors complicate the application of Earth-based models to Mars, severely impeding predictive capabilities for a broad range of phenomena, from dust lifting \cite{basu2004, kahre2006, alvarez2024} to sand fluxes \cite{vermeesch2008, bridges2012} and aeolian erosion rates \cite{arvidson1979, day2019, golombek2000}. 

First, as atmospheric density decreases, atmospheric kinematic viscosity increases, thickening the viscous sublayer. As a result, the roughness-based Reynolds number decreases (for a given roughness scale and shear velocity; Eq. \ref{eq:Rek}), increasing the likelihood of aerodynamically smooth wind events, and thus, of $z_0$ being influenced by the laminar sublayer (Eq. \ref{eq:sublayer}). However, it is unclear whether this effect is annulled by active saltation under Mars-like conditions (Eqs. \ref{eq:z0_Hf}-\ref{eq:z0_mod}). 

Second, under low-pressure conditions, two types of windblown ripples may emerge - impact ripples and drag ripples 
\cite{lapotre2016,duran2019,lapotre2021,rubanenko2022, courrech2024, yizhaq2024,alvarez2024_2}. Aeolian drag ripples are analogous to subaqueous ripples \cite{lapotre2017,lapotre2018} and aeolian dunes \cite{duran2019,rubanenko2022,courrech2024}, and result from a hydrodynamic instability that only arises under aerodynamically smooth conditions \cite{duran2019, courrech2024, yizhaq2024, alvarez2024_2}. On Mars, aeolian drag ripples are characterized by meter-scale wavelengths and decimeter-scale heights, possibly enhancing the role of form drag relative to the case of Earth's much more subdued impact ripples \cite{van1984}). 

\ca{Measurements of $z_0$ have been performed over coarse regolith and rocky surfaces on Mars before \cite{sullivan2000}, but never over active sandy rippled beds, such that} it remains unknown whether the thicker boundary layer and multiple ripple scales impact $z_0$ under active transport conditions. Here, we measure the aerodynamic roughness of rippled sand beds under varying atmospheric pressures. For the first time, we characterize the behavior of $z_0$ in low-pressure wind tunnel experiments in which two scales of ripples (impact and drag ripples) form and active saltation is present. 

\section*{Results and discussion}
\label{sec:results}

\subsection*{Aerodynamic roughness of rippled beds under active saltation}

\begin{figure}
    \centering
     \includegraphics[width=1.0\linewidth]{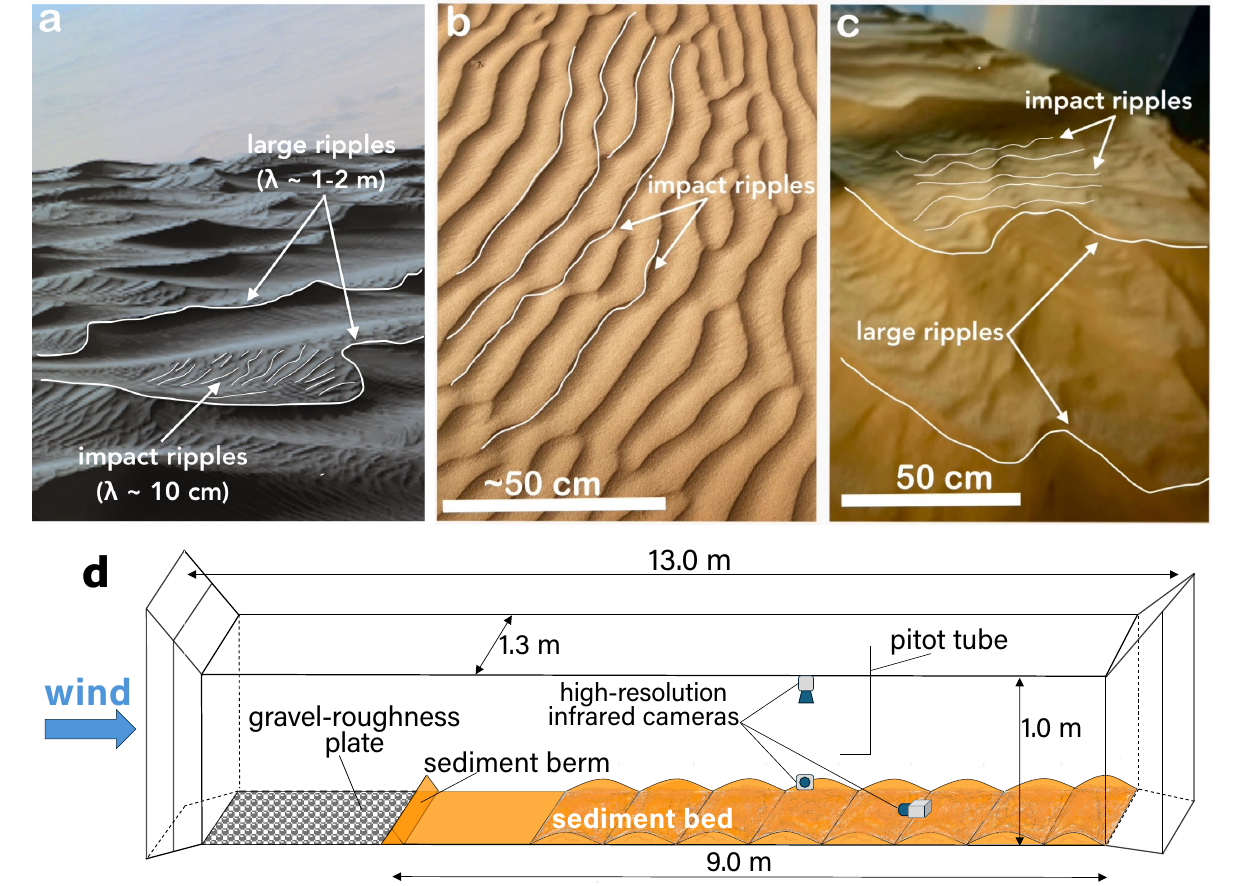}
     \caption{\textbf{Windblown ripples under varied pressure conditions and experimental setup.} \textbf{a} Impact ripples superimposed on large wind ripples atop Namib dune, Gale Crater, Mars (Curiosity rover Mastcam mosaic mcam005410, sol 1192; credit: NASA/JPL-Caltech/MSSS). \textbf{b} Impact ripples in Al Wusta, Oman. \textbf{c} Two scales of windblown ripples at equilibrium, formed in the Mars Surface Wind Tunnel (MARSWIT) at the NASA Ames Research Center under 50 mbar. \textbf{d} Schematic representation of the Mars Surface Wind Tunnel (MARSWIT) experimental setup.}
    \label{fig:fig1}
\end{figure}

\begin{figure} [h!]
    \centering
    \includegraphics[scale = 0.62]{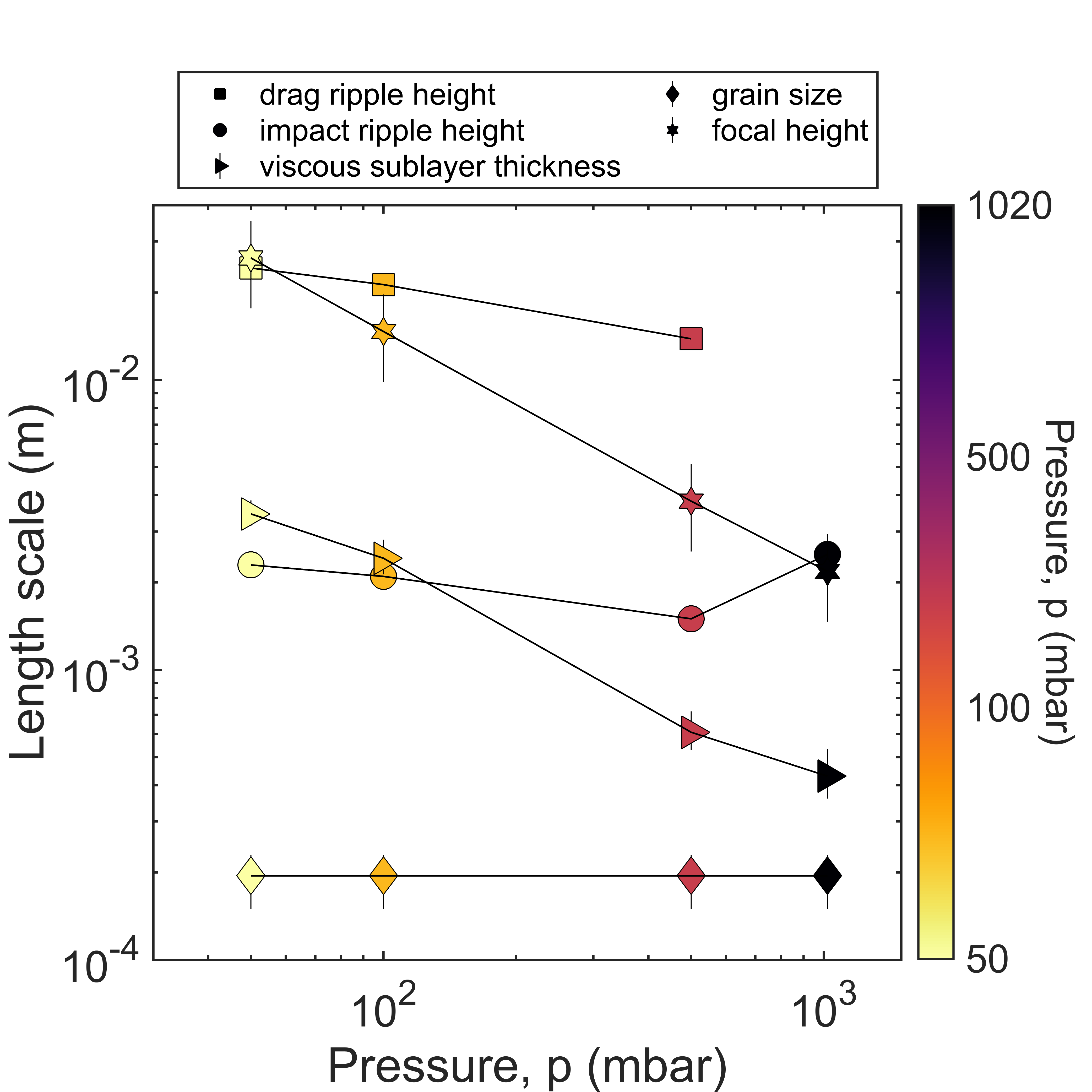}
    \caption{\textbf{Evolution of relevant length scales with atmospheric pressure during wind-tunnel experiments.} The presented scales include the grain size (reported as $d_{50}$ with error bars encompassing $d_{16}-d_{84}$), the thickness of the laminar sublayer (calculated in the absence of saltation), impact ($h_i$) and drag-ripple ($h_l$) heights, and height of the focal point, ($H_{\text{f}}$, Eq. \ref{eq:Hf}). The latter relates to the thickness of the transport layer and is calculated as described in Methods. Error bars are reported but smaller than most symbols. Source data are provided as a Source Data file.}
    \label{fig:lambda}
\end{figure}

\begin{figure}
    \centering
    \includegraphics[width=1.1\linewidth]{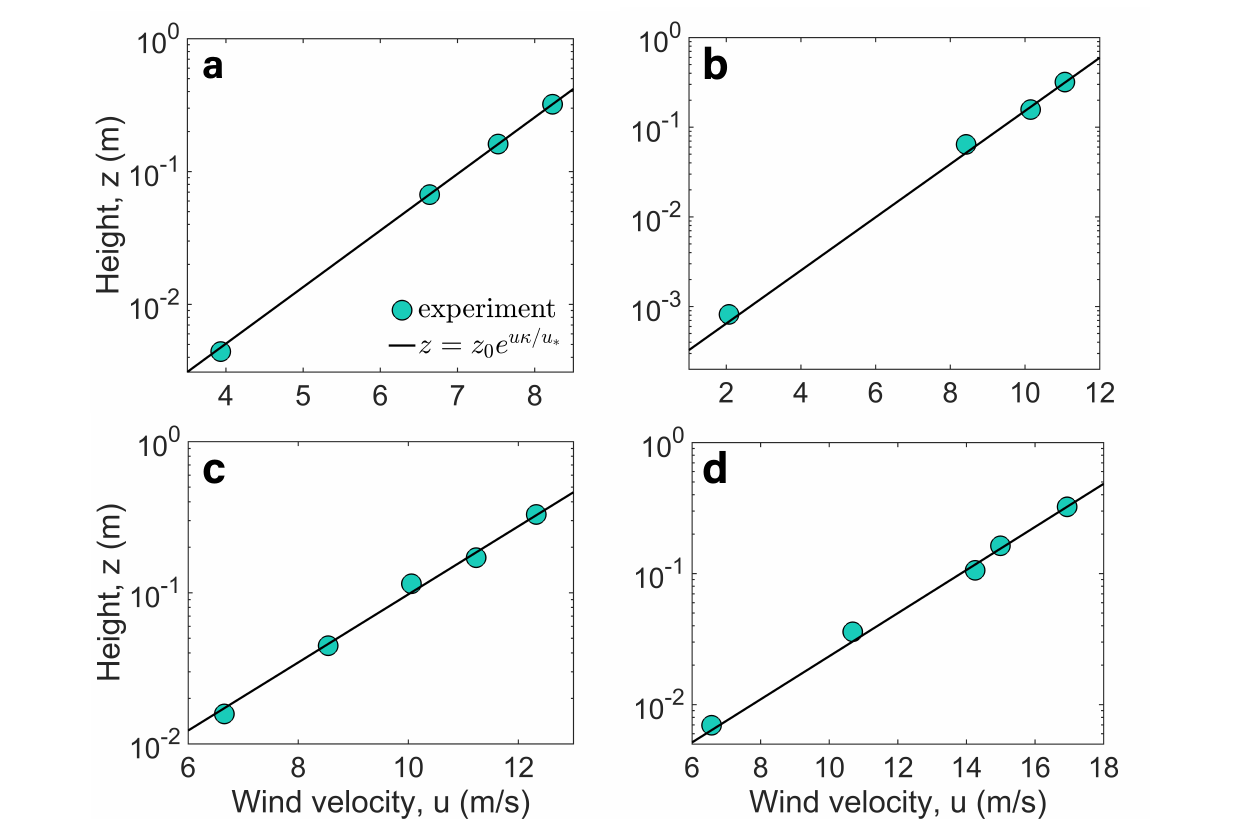}
    \caption{\textbf{Measured vertical wind velocity profiles at varying pressures with corresponding fits to the logarithmic law (Eq. \ref{eq:log_law}).} Profiles are shown for atmospheric pressures of \textbf{a} 1,020 mbar, \textbf{b} 500 mbar, \textbf{c} 100 mbar, and \textbf{d} 50 mbar.}
    \label{fig:fits}
\end{figure}

\begin{figure}
    \centering
    \includegraphics[scale = 0.51]{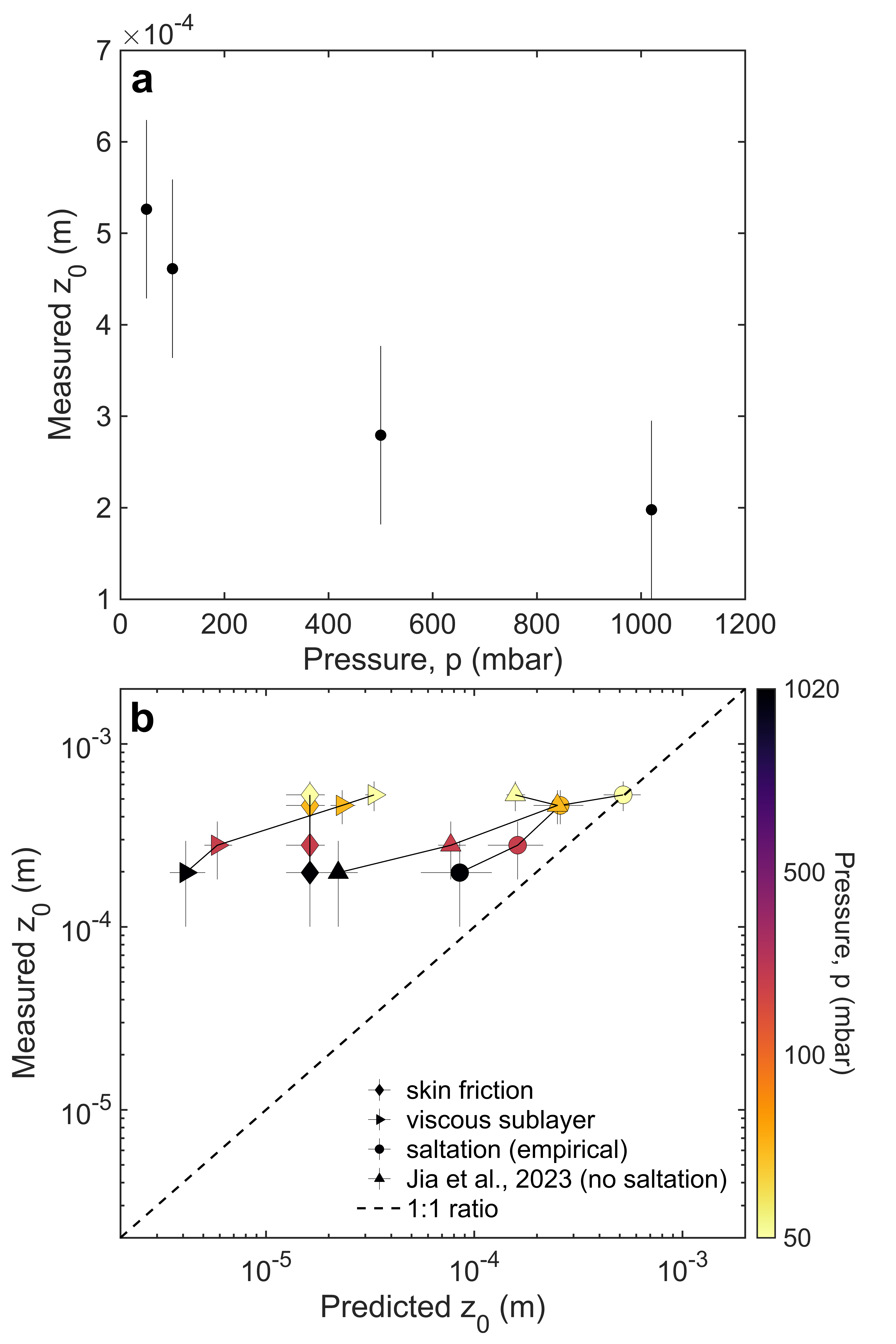}
    \caption{\textbf{Aerodynamic roughness length.} \textbf{a} Measured roughness length for different pressures. \textbf{b} Comparison of measured and modeled aerodynamic roughness length scales under Earth-to-Mars pressure conditions. Modeled values are estimated from the thickness of the viscous sublayer (Eq. \ref{eq:sublayer}), skin friction (Eq. \ref{eq:z0_rough}), saltation-layer thickness (Eq. \ref{eq:z0_mod}), and compound roughness from skin friction and form drag from all bedform scales present \cite{jia2023} (Jia et al., 2023, no saltation). Source data are provided as a Source Data file.}
    \label{fig:z0_model}
\end{figure}

Experiments were conducted over beds of loose sand-sized proxy material ($\sim$195 $\mu$m crushed nutshells) at freestream wind velocities exceeding the corresponding freestream threshold for the onset of saltation by $\approx 20\%$. Beds were allowed to develop bedforms and equilibrate (Figs. \ref{fig:fig1}c and \ref{fig:lambda}) \cite{alvarez2024_2} before vertical wind profiles were measured using a variable-height pitot tube at a frequency of 0.3 Hz (Fig. \ref{fig:fig1}d). Measurements were performed at 4-5 different heights above the bed, and averaged over 45-60 seconds at each height \cite{alvarez2024_2} (Methods). Wind profiles were measured at 1,020, 500, 100, and 50 mbar, and $z_0$ was derived from those measurements by finding the best fit $\left(u_*,z_0\right)$ values using the logarithmic law (Eq. \ref{eq:log_law}; Figs. \ref{fig:fits} and \ref{fig:z0_model}; Methods). \ca{These experiments, which capture the correct physical regimes despite higher absolute pressures than on Mars, allow us to extrapolate observations to Mars-like surface conditions (Supplementary Text S1 and Supplementary Table S1).}

%\dummyfigure{fig:fig1}

Decimeter-scale impact ripples formed at all pressures \cite{andreotti2021}, whereas larger, drag ripples only formed at pressures less than terrestrial (Fig. \ref{fig:lambda}), as predicted by drag-ripple theory \cite{duran2019,courrech2024, alvarez2024_2}. The size of drag ripples increased with decreasing atmospheric pressure, consistent with observations on Mars \cite{lapotre2016,rubanenko2022,vaz2023} (Fig. \ref{fig:fig1}a). Similarly, the thickness of the viscous sublayer in the absence of saltation (calculated as $11.6 \frac{\nu}{u_*}$) is expected to increase by about one order of magnitude from ambient terrestrial pressure to 50 mbar (Fig. \ref{fig:lambda}). 

%\dummyfigure{fig:lambda}

%\dummyfigure{fig:fits}

Like several of the relevant length scales (Fig. \ref{fig:lambda}), measured $z_0$ values increase with decreasing atmospheric pressure, from $\approx 2 \times 10^{-4}$ m at 1,020 mbar to over $5 \times 10^{-4}$ m at 50 mbar under our experimental conditions (Fig. \ref{fig:z0_model}a). Similarly, predicted values of $z_0$ from viscous sublayer thickness and saltation increase with decreasing pressure (Fig. \ref{fig:z0_model}b) due to decreasing atmospheric density and increasing threshold of motion, respectively (Supplementary Figure S3). 

Derived values of $z_0$ are compared against model predictions (Fig. \ref{fig:z0_model}b; Eqs. \ref{eq:sublayer}-\ref{eq:z0_mod}; and model of Ref. \cite{jia2023}; Methods). The large discrepancy between measured $z_0$ and that expected from the viscous sublayer model is consistent with either aerodynamically rough conditions (at 1,020 mbar) or disruption of the viscous sublayer by saltation at all pressures. At 1,020 mbar, the model of Ref. \cite{jia2023}, which predicts the compounded effect of skin friction and form drag from impact ripples, is dominated by skin friction and underpredicts the measured $z_0$ value, confirming that saltation (Eq. \ref{eq:z0_mod}) is the dominant control on $z_0$ under Earth-like conditions. At all pressures, observed $z_0$ are most closely matched by the empirical saltation model (Eq. \ref{eq:z0_mod}) and the compound form drag model of Ref. \cite{jia2023}. However, both models consistently underpredict $z_0$, suggesting that both saltation and form drag contribute to overall aerodynamic roughness. 

\begin{figure}
    \centering
    \includegraphics[scale=0.09]{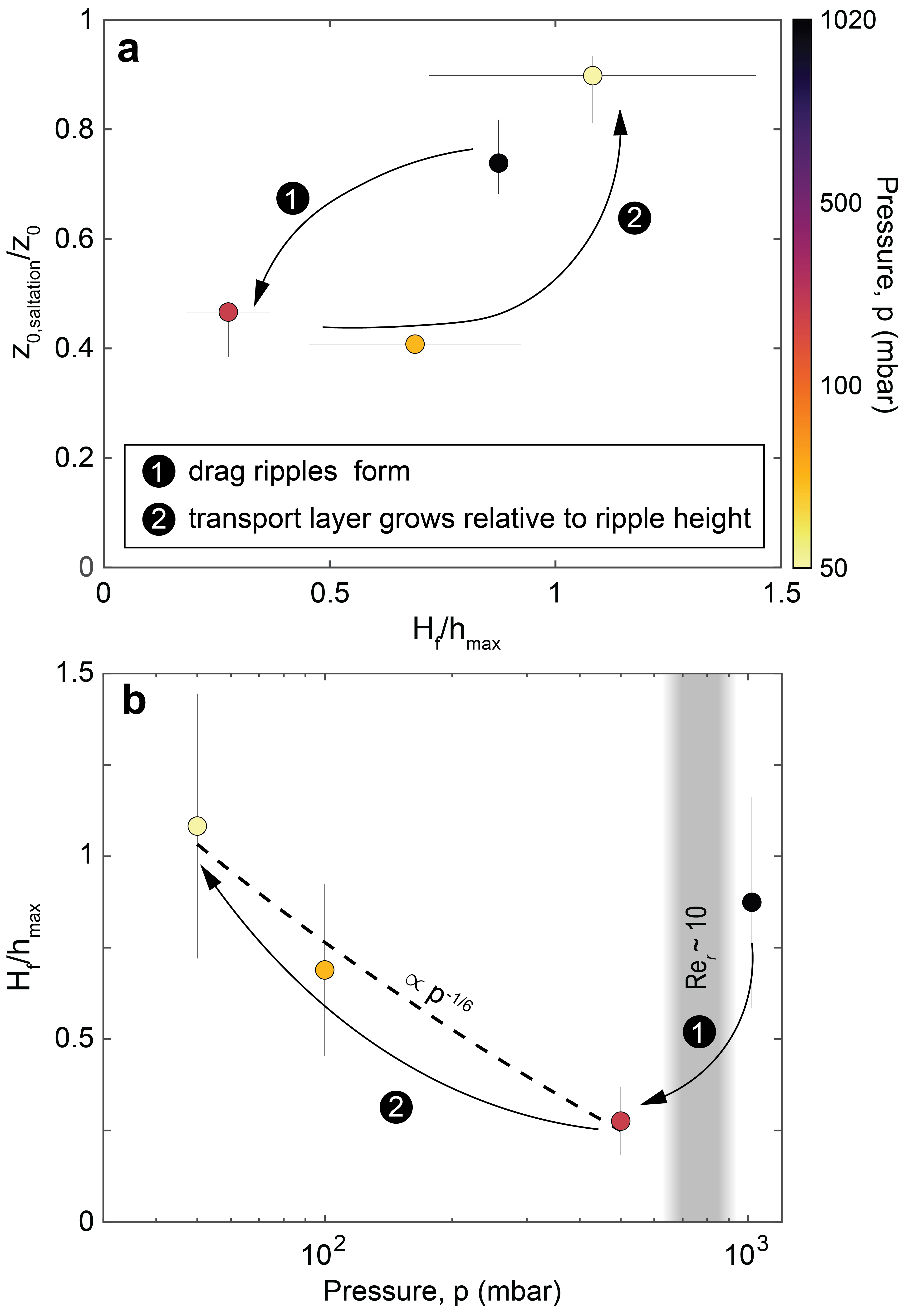}
    \caption{\textbf{Influence of saltation on aerodynamic roughness under varied atmospheric pressures.}  \textbf{a} Ratio of aerodynamic roughness induced by saltation, $z_{0\text{,saltation}}$ (modeled) and $z_0$ (experimental measurements) as a function of the focal height, $H_{\text{f}}$, normalized by the height of the largest ripples present, $h_{\text{max}}$, at each pressure. \textbf{b} Ratio of focal height to maximum bedform height, $H_{\text{f}}/h_{\text{max}}$, as a function of pressure. Shaded area indicates the boundary between aerodynamically smooth and aerodynamically rough conditions (defined by $Re_r$ $\sim$ 10). Source data are provided as a Source Data file.}
    \label{fig:Hf_h}
\end{figure}

Next, we investigate the relative influence of saltation and form drag on $z_0$. To estimate the contribution of saltation to $z_0$ based on our experimental measurements, we use the model of Ref. \cite{jia2023} in an inverse fashion (Methods). Specifically, we adjust the value of an effective grain size in this model such that, given the observed bedform dimensions at each pressure, it produces the correct, observed value of $z_0$. Assuming that skin friction is negligible relative to other roughness contributors under active saltation (Fig. \ref{fig:z0_model}b), the effective grain size parameter, instead of grain size, represents a characteristic length scale associated with the saltation layer, from which $z_{0,\text{saltation}}$ can be calculated (Fig. \ref{fig:Hf_h}; Methods; \cite{jia2023}).

%\dummyfigure{fig:z0_model}

%\dummyfigure{fig:Hf_h}

The relative influence of saltation on total aerodynamic roughness, $z_{0\text{,saltation}}/z_0$, varies in a nonlinear fashion with atmospheric pressure (Fig. \ref{fig:Hf_h}a). Specifically, we find that $z_{0\text{,saltation}}$ is about 80$\%$ of $z_0$ at 1,020 mbar, when only subdued impact ripples are present. As pressure decreases and drag ripples form, the contribution of saltation to $z_0$ drops to $\approx 45 \%$ at 500 mbar, and increases again progressively under our experimental conditions as pressure decreases, to reach about $\approx 90 \%$ at 50 mbar (Fig. \ref{fig:Hf_h}b).

%\section*{Discussion}
%\label{sec:discussion}

\subsection*{Focal point at low pressure}

Although the existence of a focal point is well established under terrestrial conditions (see, e.g., Refs. \cite{bagnold1936, ungar1987, duran2011, ho2014, valance2015}), how viscosity may influence its existence, and if it exists, its behavior at low atmospheric pressures remain unknown. Here, we test its existence and behavior by (i) inverting for $H_\text{f}$ and $U_\text{f}$ from $z_{0,\text{saltation}}$ using Eq. \ref{eq:z0_Hf} and (ii) comparing $H_\text{f}$ and $U_\text{f}$ values with \ca{the models of Refs. \cite{andreotti2004} and \cite{pahtz2021}}. 

Based on dimensional analysis, theory, and empirical observations, Refs. \cite{andreotti2004} and \cite{pahtz2021} proposed formulations for focal height and velocity that incorporate the influence of atmospheric density and viscosity (Methods). \ca{Each model contains two constant coefficients, $\alpha$ and $\beta$ (Methods). \ca{For both models,} best-fit values of $\alpha$ and $\beta$ were inverted at all pressure conditions by minimizing the difference between the estimated $z_{0,\text{saltation}}$ and those modeled using the formulations of Refs. \cite{andreotti2004} and \cite{pahtz2021} (Methods).}

\ca{Using the model of Ref. \cite{andreotti2004},} we find that (i) inverted values of $\alpha$ and $\beta$ are roughly constant across the investigated pressure range, and (ii) inverted $\beta$ values closely match that estimated by Ref. \cite{andreotti2004} at Earth pressure (Supplementary Figure S7). Inverted values of $\alpha$ are of the same order of magnitude as, but differ subtly from, those of Ref. \cite{andreotti2004} at Earth pressure (Supplementary Figure S7). This small difference is attributed to the presence of impact ripples in the experiments of Ref. \cite{rasmussen1996} (and the form drag they imparted on winds), whereas any contributions from form drag is removed from measured $z_0$ values here (i.e., $\alpha$ and $\beta$ are estimated from $z_{0,\text{saltation}}$). These findings suggest that a focal point does exist under all investigated pressures, and that the formulation of Ref. \cite{andreotti2004} appropriately describes the dependency of $H_\text{f}$ and $U_\text{f}$ on atmospheric density and viscosity.

\ca{Whereas sets of $\alpha$ and $\beta$ values that yield equally good fits to the data could be retrieved from both models, we find that $\beta$ cannot be reasonably assumed to be constant over the range of investigated pressures under the model of Ref. \cite{pahtz2021}, and that $\beta \propto \rho^{-\frac{1}{3}}$ instead (Supplementary Text S2). Thus, the model of Ref. \cite{andreotti2004} appears to capture the dependency of $H_\text{f}$ on $\rho$ more adequately, and we adopt that formulation for extrapolations to Mars pressure.}

\subsection*{Relative influence of saltation and form drag on aerodynamic roughness}\label{sub_salt}

Following the model of Ref. \cite{andreotti2004}, $H_\text{f} \propto \rho^{-\frac{5}{6}}$ (Eqs. \ref{eq:Uf}-\ref{eq:Hf}; Methods). Furthermore, the wavelength of drag ripples is proportional to $\rho^{-\frac{2}{3}}$ \cite{lapotre2016, lapotre2017, lapotre2018, lapotre2021, rubanenko2022, alvarez2024_2}. Thus, under low atmospheric pressures (such that Re$_r \lesssim 10$), $h_{\text{max}} \propto \rho^{-\frac{2}{3}}$ (for a given bedform aspect ratio). As a result, 
\begin{equation}
\frac{H_\text{f}}{h_{\text{max}}} \propto \rho^{-\frac{1}{6}} \propto p^{-\frac{1}{6}}
\label{eq:Hf_h_andreotti}
\end{equation}
at a constant temperature. \ca{This prediction, which relies on $\beta$ being independent of atmospheric density, is well matched by the data (R$^2$ = 0.98; Fig. \ref{fig:Hf_h}b).}

At Earth pressure, drag ripples cannot form and $\frac{H_\text{f}}{h_{\text{max}}} \approx \ca{90} \%$. As pressure decreases, drag ripples form, protruding higher into the flow ($\frac{H_\text{f}}{h_{\text{max}}} \approx \ca{25} \%$ at 500 mbar; Fig. \ref{fig:Hf_h}b) enhancing the contribution of form drag to total aerodynamic roughness (Fig. \ref{fig:Hf_h}a). As pressure further decreased in our experiments, the saltation layer thickened faster than drag ripples grew, such that $\frac{H_\text{f}}{h_{\text{max}}}$ increased with decreasing pressure.

Whereas $H_{\text{f}}$ increases with decreasing pressure, so does $U_{\text{f}}$, such that the behavior of $z_{0,\text{saltation}}$ may vary depending on specific transport conditions (Eq. \ref{eq:z0_Hf}). In our experiments, wind speed was increased with decreasing pressure such that the freestream wind velocity remained $\approx 20\%$ above the saltation threshold (Methods). As a result, $z_{0,\text{saltation}}$ overall increased with decreasing pressure, enhancing the contribution of saltation to total aerodynamic roughness despite the formation of taller and taller drag ripples (Fig. \ref{fig:Hf_h}). At 50 mbar, $\frac{H_\text{f}}{h_{\text{max}}} \approx \ca{110} \%$ and $\frac{z_{0\text{,saltation}}}{z_0} \approx 90 \%$.

\begin{figure}[th]
    \centering
    \includegraphics[width=0.7\linewidth]{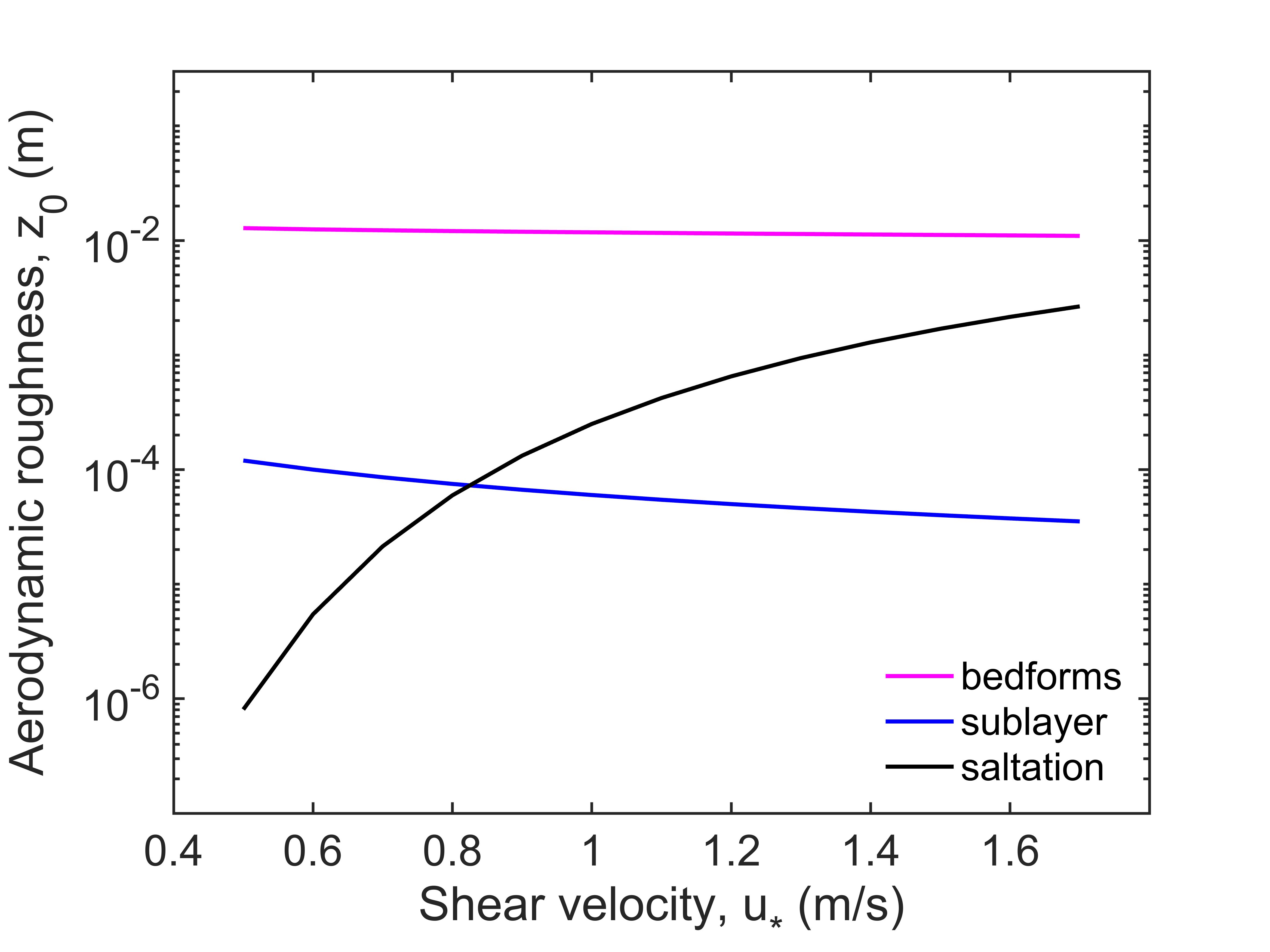}
    \caption{\textbf{Aerodynamic roughness, $z_0,$ as a function of shear velocity, $u_*$, on Mars.} Aerodynamic roughness due to compounded skin friction and bedforms (bedforms) was calculated using the Jia et al. (2023) model \cite{jia2023}. Aerodynamic roughness due to saltation (saltation) was estimated using the Andreotti (2004) model \cite{andreotti2004} and Eq. \ref{eq:z0_Hf}, along with calibrated values of $\alpha$ and $\beta$ (Eqs. \ref{eq:Uf}-\ref{eq:Hf}). Source data are provided as a Source Data file.} 
    \label{fig:z0_mars}
\end{figure}

%\dummyfigure{fig:z0_mars}

\subsection*{Implications for Mars}

Our wind-tunnel experiments demonstrate that the roles of transport and bedforms on $z_0$ do not follow monotonic trends with atmospheric pressure, but instead are affected by interactions between bedform development and transport layer dynamics. \ca{Extrapolations to Mars-like conditions thus require careful considerations of these factors.}

\ca{Dimensional analysis indicates that our experimental conditions capture the physical phenomenology expected on Mars (aerodynamically smooth beds conducive to drag-ripple formation with sand transport in saltation) for neutral atmospheric conditions (Supplementary Text S1 and Supplementary Table S1). Restated, experimental conditions reproduce the correct physical regimes despite different pressure conditions. The assumption of neutral atmospheric conditions is implicit to Eq. 1. Theoretical \cite{monin1954}, experimental \cite{white1991martian}, and in situ \cite{sullivan2000} evidence indicate that thermally induced variations in atmospheric conditions (e.g., from diurnal temperature swings) results in minor deviations in $z_0$ that are well within the uncertainty of our measurements (Supplementary Text S1 and Supplementary Table S1)}.

Based on our validation and calibration of Eqs. \ref{eq:Uf}-\ref{eq:Hf}, we can estimate $z_{\text{0,saltation}}$ on Mars. In parallel, we estimate the combined contributions of skin friction and form drag from all bedform scales ($z_{\text{0,bedforms}}$) using the Jia et al. (2023) model \cite{jia2023} assuming grain and bedform sizes as measured at Gale crater (grain size, $d = 125~\mu$m \cite{weitz}; impact-ripple wavelength, $\lambda_i$ = 8 cm; drag-ripple wavelength, $\lambda_l$ = 2 m; bedform height-to-wavelength ratios of 0.05 \cite{lapotre2016,lapotre2018}). Typical values for Mars were used for all other parameters (e.g., $\rho$ = 0.020 kg/m$^3$, $\rho_s = 2,900$ kg/m$^3$, $g$ = 3.71 m/s$^2$). Both aerodynamic length scales are also compared with $z_{0\text{,sublayer}}$ (Eq. \ref{eq:sublayer}).

We find that $z_{\text{0,bedforms}} \approx 1$ cm, which is consistently 1–2 orders of magnitude larger than $z_{\text{0,sublayer}}$ and $z_{\text{0,saltation}}$ across a range of plausible wind shear velocities (Fig. \ref{fig:z0_mars}), i.e., it is 1-2 orders of magnitude larger than over a flat bed. Thus, $z_0$ is expected to be dominated by form drag over rippled beds on Mars regardless of whether transport is ongoing. \caa{This extrapolation implicitly relies on the assumption that flux saturation was reached within our wind-tunnel experiments. Estimates of the sediment transport saturation length under our experimental conditions are lower than the length of the wind-tunnel's test section (Methods), though the true distance required to reach saturation from a zero flux can be several times larger than the saturation length \cite{selmani2018}. Measurements of impact-ripple migration speeds in the experiments (\cite{alvarez2024_2}; Supplementary Figure S8) and our estimates of focal-height parameters (Supplementary Figure S7) compare favorably with models that were developed to describe fully saturated conditions \cite{andreotti2004,duran2011}, suggesting that near-saturation conditions were likely reached. However, these models were themselves calibrated using empirical constants that could possibly obscure the signature of non-saturated transport conditions. If transport was substantially under-saturated in our experiments, the above extrapolations to Mars would underpredict $z_{0\text{,saltation}}$, rendering the contribution of saltation to $z_0$ comparable to that of form drag from bedforms.}

Finally, \caa{we note that} calculating $r$ from $\text{max}\left(z_{\text{0,sublayer}},z_{\text{0,saltation}}\right)$ \cite{jia2023}, we find that $\text{Re}_r < 10$ - even during active transport - over flat beds on Mars. Specifically, we find that wind shear velocities, $u_* > 1$ m/s are required for $\text{Re}_r$ to exceed 10. Extrapolated to the height of the wind sensor onboard NASA's Curiosity rover (1.5 m above ground), such wind shear velocities correspond to wind speeds of $>$ 22.5 m/s (Eq. \ref{eq:log_law}), or stronger than over 99.9$\%$ of winds measured by Curiosity at Gale Crater \cite{viudez2019}. This result is significant because it disproves the notion that saltation on Mars would inhibit the formation of drag ripples by promoting aerodynamically rough flat-bed conditions.

\section*{Methods}
\label{sec:methods}

\subsection*{Experimental conditions}
\label{subsec:experiments}

Experiments were conducted in the Mars Surface Wind Tunnel (MARSWIT) at the Planetary Aeolian Laboratory, NASA Ames Research Center, California. The MARSWIT operates within a pressure chamber capable of simulating Earth-to-Mars pressure conditions (approximately 1 bar to 5 mbar). This 13-m long, open-circuit, atmospheric boundary layer wind tunnel can achieve maximum wind speeds of 100 m/s at 5 mbar using air injectors. LabView data acquisition software is utilized for operations, measuring and providing data on chamber temperature, air density, pressure, freestream and vertical differential pressures, pitot tube vertical position, relative humidity, and corresponding wind speeds. \caa{The pitot tube was placed at the downwind end of the test section. The test section was approximately 6.5 m long and 1.3 m wide (Fig. \ref{fig:fig1}).}

For each experiment, pressure, air density, and temperature were recorded, and the dynamic viscosity of air was assumed to be 1.81 $\times$ 10$^{-5}$ Pa$\cdot$s. Crushed nutshells ($\rho_s$ = 1,300 kg/m$^3$) served as sediment with a median particle size ($d_{50}$) of 195 $\mu$m. Grain-size distributions were determined using a Retsch Camsizer grain-size analyzer. Nutshells were used not to simulate Mars' lower gravitational acceleration (g = 3.71 m/s$^2$) on saltation trajectories but to reduce the time required to generate bedforms under our lowest experimental pressure conditions (50 and 100 mbar) \cite{alvarez2024_2}. \caa{A manually smoothed, flat sediment bed was initially prepared inside the wind tunnel before each experiment}, ensuring uniform sediment distribution across the test section (Supplementary Figure S2). The bed had a thickness of 3 cm across the wind tunnel, and a sediment berm was placed at the upwind end of the bed to act as a continuous sediment source. 

Two series of experiments were performed. In the first series, we determined the onset of saltation over flat beds by progressively increasing the wind speed until the bed became fully mobile \cite{swann2020}. This setup allowed for the determination of the threshold freestream ($U_{\infty}$) and shear ($u_{*t}$) velocities over a flat bed at pressures of 50, 100, 500, and 1,020 mbar (Supplementary Figure S3). In the second series, the wind velocity was set $\approx 20\%$ above the threshold freestream velocity, and maintained for several minutes to allow the sediment bed to evolve naturally from a flat to a rippled configuration. Once the ripples reached equilibrium, wind velocity measurements were performed at pressures of 50, 100, 500, and 1,020 mbar. In the latter series of experiments, multiple runs (pumpdowns) were sometimes necessary at a given atmospheric pressure to replenish the berm and maintain depositional conditions while the bedforms reached equilibrium. For the 1,020 mbar and 500 mbar experiments, a single pumpdown sufficed, while the 100 mbar experiments required two pumpdowns, and the 50 mbar experiments required three. \caa{Transport saturation lengths under our experimental conditions were estimated two different ways using the formulations in Refs. \cite{duran2019, yizhaq2024}, and were found to always be shorter to comparable to the length of the test section ($\sim$ 6.5 m), even under our lowest pressure condition ($\sim$ 5.5-6 m at 50 mbar).}

%\subsection{Shear velocity ($u_*$) and threshold shear velocity ($u_{*t}$)}
\label{sec:velocity}

\subsection*{Ripple dimensions}
\label{subsec:ripples}

At each specific pressure, the evolution of the bedforms was continuously recorded with high-resolution infrared cameras at 30 fps, which provided top-down, side, and windward views of the test section (Fig. \ref{fig:fig1}). This setup enabled precise measurement of the wavelength and migration rate (celerity) of incipient ripples and their final stabilized states. Each test concluded once the bedforms ceased growing, indicating equilibrium (Fig. \ref{fig:fig1}c). At the end of each experiment, after the winds stopped, the pressure chamber was reset to ambient pressure. This allowed for additional documentation of the final bed state through three-dimensional (3D) scans and high-resolution color imaging. 3D scans were acquired using LiDAR sensors built into an iPad Pro 4th generation \cite{luetzenburg2021} with Pix4Dcatch software, and the data were processed with MeshLab 2023.12 (Supplementary Figure S4). These scans allowed us to document the dimensions of equilibrium bedforms along downwind-oriented topographic profiles, including crest-to-crest wavelength and trough-to-peak height, under consistent lighting conditions. The 3D scans enabled the accurate calculation of the height and wavelength of drag (large) ripples, as well as the wavelength of the impact (small) ripples. The height of impact ripples was calculated as $5\%$ of measured ripple wavelengths \cite{ellwood1975}. Calculated impact ripple heights range from approximately 1 to 3 mm at all pressures, in good agreement with ripple heights as determined from the length of ripple shadows at 50 mbar \cite{alvarez2024_2}(Supplementary Figure S5). Bedform dimensions are reported in Supplementary Table S2 for each pressure condition.

\subsection*{Measurements of aerodynamic roughness length, $z_0$}
\label{subsec: z0}

Wind speed measurements were conducted using a pitot tube positioned at varying heights above the sediment bed. Wind velocity data were collected at four to five different vertical positions - at elevations of 4, 33, 103, 160, and 320 mm above the initial surface elevation - for 45 to 60 seconds at each position, with a sampling frequency of 0.3 Hz. The elevations were automatically set using a dedicated routine in LabView, with a precision of $\pm$ 0.05 mm (Supplementary Figure S1). The control system can maintain a constant wind speed with variations of less than 1$\%$ of the set value. In the same wind tunnel facility, Ref \cite{swann2020} performed wind measurements at nearly the same elevations above the bed to determine the threshold shear velocity of grains similar to those in this study. Changes in bed elevation were continuously monitored using a high-resolution infrared camera positioned for a side-view perspective. The initial height of the bed was checked relative to the position of the pitot tube and confirmed using a digital caliper. For experiments that required more than one pumpdown, i.e., at 100 and 50 mbar, the height of the bed was also monitored with a digital caliper between pumpdowns as the ripples started to form, and a new reference level was recorded.

The wind speed measurements were used to calculate the threshold shear velocity $u_{*t}$ (flat-bed series) or shear velocity $u_*$ (equilibrium-rippled bed) using the law of the wall (Eq. \ref{eq:log_law}). \ca{To that end, we used MATLAB's built-in Nonlinear Least Squares function. The method works by finding the parameters of a nonlinear model that minimize the sum of the squared differences (residuals) between the observed data and the model predictions (Eq. \ref{eq:log_law})}. Instantaneous measurements provided sets of 135 to 180 data points for each elevation. The time between measurements at each elevation was 45 to 60 seconds, allowing to average wind measurements over turbulent fluctuations \cite{stull1988}. Only fits with $R^2$ values larger than 0.98 were considered, \ca{which resulted in the exclusion of 8–13$\%$ of the dataset. Remaining data} were averaged to determine $u_*$ and $z_0$ at each specific pressure (Fig. \ref{fig:fits}). As measurements for each elevation were repeated 2 or 3 times, an additional averaging process was performed to obtain the final values of $u_*$ and $z_0$. Uncertainty in $u_*$, and $u_{*t}$ was calculated as $\pm$ one standard deviation, and in $z_0$ as $\pm$ $d/2$. Values and uncertainties of $u_{*t}$, $u_*$, and $z_0$ for each pressure condition are reported in Supplementary Table S2.

\subsection*{Scaling relationship for the focal point}
\label{subsec:focal}
Ref. \cite{andreotti2004}, using experimental surface roughness data from Ref. \cite{rasmussen1996}, showed that under terrestrial conditions the focal velocity and focal height can be reasonably well predicted by 

\begin{equation}
    U_{\text{f}} \approx \alpha \sqrt{\frac{\rho_s}{\rho}gd},
    \label{eq:Uf}
\end{equation}
\noindent and
\begin{equation}
     H_{\text{f}}\approx \beta_A U_{\text{f}} \left(\frac{\nu} {g^2}\right)^{1/3},
     \label{eq:Hf}
\end{equation}

\noindent where $\alpha$ and $\beta_A$ are constants, $\rho_s$ and $\rho$ are the sediment and atmospheric densities, respectively, $g$ is the acceleration of gravity, and $d$ is grain diameter. 

\ca{In turn, Ref. \cite{pahtz2021} proposed a different formulation for the focal height,}

 \begin{equation}
    H_{\text{f}} \approx \beta_{P+T} \left(\frac{\nu^2}{g}\right)^{1/3},
    \label{eq:Hf_pahtz}
 \end{equation}
\noindent \ca{where $\beta_{P+T}$ is a constant.}

\ca{Using an optimization subroutine, we minimized the difference between each of the two models and $z_{\text{0,saltation}}$. This approach allowed us to iteratively solve for sets of $\alpha$ and $\beta$ parameters across all pressure conditions}. 

\ca{As reported in the main text, both models yield satisfying fits to the data, but $\beta_{P+T}$ is not well approximated by a constant over the range of investigated pressures. Eqs. \ref{eq:Uf} and \ref{eq:Hf} were thus adopted, and the average of inverted $\alpha$ and $\beta_A$ values were used in our analysis (1.707 and 0.163, respectively; Fig. \ref{fig:lambda}). Uncertainty in $\alpha$ and $\beta$ was calculated as $\pm$ one standard deviation.}

\subsection*{Model for aerodynamic roughness induced by multiscale topography}
\label{subsec:model}

Experimentally derived $z_0$ values are compared to the theoretical model of Ref. \cite{jia2023}, which predicts the influence of multiscale topography (grains of size $d$, and small and large ripples characterized by wavelengths, $\lambda$, and amplitudes, $\zeta$). In this model, $z_0$ is calculated for three scales by considering the roughness induced by grains on small ripples, the roughness induced by small ripples on large ripples, and the roughness induced by large ripples themselves. A complete description of the model can be found in Ref. \cite{jia2023}.

Our calculations of $u_*$, the height and wavelength of both small and large ripples ($h_{i,l}$ and $\lambda_{i,l}$), as well as the experimental data on kinematic viscosity of air, $\nu$, and the median grain size ($d$ =195 $\mu$m), were used as input to the model. The effective roughness is calculated for a specific bedform aspect ratio, 

\begin{equation}
    k \zeta = \frac{2 \pi}{\lambda_{i,l}} \, \frac{h_{i,l}}{2},
    \label{eq:aspect_ratio}
\end{equation}

\noindent where $k$ is the bedform's wavenumber. Compound roughness values from skin friction and all bedform scales are reported for each pressure condition in Supplementary Table S2.

The model of Ref. \cite{jia2023} underestimates measured $z_0$ values (Fig. \ref{fig:z0_model}) because, while it accounts for form drag from two scales of bedforms, it does not incorporate the influence of saltation. Thus, comparing measured and modeled values of $z_0$ allows to quantify the relative influence of saltation ($z_{0\text{,saltation}}$). The model of Ref. \cite{jia2023} was used to that end as described in the Results and discussion section. To further validate this approach, the inverted value of $z_{0\text{,saltation}}$ at 1,020 mbar was compared to predictions from numerical simulations \cite{duran2011} (Supplementary Figure S6).

\section*{Data availability}

The experimental parameters and derived measurements generated in this study are provided in the Supplementary Information file. Source data are provided with this paper.

\section*{Code availability}

Code to reproduce the data shown in the figures is available in the Zenodo database under accession code https://doi.org/10.5281/zenodo.14970357

%\bigskip

%\newpage

%\bibliography{marswit}% common bib file

\bigskip

\newpage
\section*{Acknowledgements} The authors thank Andrew Gunn, James Ken Smith, and Farid Haddad for help with the experimental setup and facility, \ca{and Bruno Andreotti for insightful discussions.}

This work was partly funded by NASA under Grant No. 80NSSC20K0145 to M.G.A.L., C.S., and R.C.E.; P.J. was supported by the National Natural Science Foundation of China (NSFC12102114). 

\section*{Author contributions} C.A.A. and M.G.A.L. designed the experiments, and C.A.A. performed the experiments. C.A.A., M.G.A.L., C.S., R.C.E., P.J., and P.C. contributed to the conceptualization and formal analysis. C.A.A. and M.G.A.L. wrote the paper. P.J. ran the multi-scale topography model. All authors discussed the results and implications and commented on the paper at all stages.

\section*{Competing interests}
The authors declare no competing interests.

\bigskip

\newpage
%------------------------------------------------------------------
\section*{Appendix S1: Applicability of experimental measurements to Mars situation}

We use dimensional analysis to verify that the range of investigated pressure conditions is sufficient for extrapolations down to Mars pressure. The length scale $z_0$ over active rippled beds is determined by $n = 6$ parameters: grains diameter, $d$ (which controls the roughness height, $r \simeq 2.5d$), bedform size, $\lambda$, shear velocities, $u_*$, kinematic viscosity, $\nu$, and sediment and fluid densities, $\rho_s$ and $\rho$. Together, these $6$ parameters encompass $k = 3$ dimensions (mass, length, and time). Thus, according to the Buckingham $\Pi$ theorem, any set of $n-k = 3$ independent dimensionless numbers suffice to describe the physical system. 

We use a set of two Reynolds numbers (associated with roughness and ripple length scales) and a Shields number. The roughness-based Reynolds number, ${\rm Re}_r$, controls the aerodynamic roughness of the bed, with a transition from aerodynamically smooth to aerodynamically rough conditions at ${\rm Re}_r \simeq 10$. The bedform-based Reynolds, ${\rm Re}_\lambda$, arises in drag-ripple theory \cite{abrams1985, Frederick1988, duran2019} and separates ripples (${\rm Re}_\lambda < 10^4$) from dunes (${\rm Re}_\lambda > 10^4$). Finally, the Shields number, $\Theta$, characterizes flow strength relative to sediment-stabilizing forces, and dictates the mode of sediment transport. Values of $\Theta > \Theta_t$ (threshold) is required for transport in saltation to occur, whereas values of $\Theta \gtrsim 0.1$ ($\gg \Theta_t$) indicate transport in suspension.

Table S1 summarizes the values of all three dimensionless quantities under each of our experimental conditions and as estimated on Mars. At 50 mbar, all three dimensionless numbers fall within the same regimes as on Mars -- aerodynamically smooth beds conducive to drag-ripple formation and sand transport in saltation. Therefore, our lowest experimental pressure effectively captures the physics at play on Mars, despite an order of magnitude difference in atmospheric pressure.

%------------------------------------------------------------------
\section*{Appendix S2: Focal height model}

The model of P\"athz and Tholen \cite{pahtz2021} predicts that the focal height $H_{\rm f}$  is proportional to $\rho^{-2/3}$, such that $H_{\rm f}/h_{\rm max}$ (where $h_{\rm max}$ is the height of the largest ripples present at a given pressure) is independent of $\rho$ -- as long as $\beta$ is constant. Yet, calculated values of $H_{\rm f}/h_{\rm max}$ are poorly fit to a constant $\rho$ ($R^2= 0.82$), indicating that $\beta \propto \rho^{-1/3}$ instead ($R^2 = 0.99$). Thus, the model of Andreotti (2004) \cite{andreotti2004} better captures the full dependency of $H_{\rm f}$ on $\rho$, justifying its adoption for extrapolations to Mars-like pressure conditions.

%------------------------------------------------------------------
\newpage
\renewcommand\thetable{S\arabic{table}}
\setcounter{table}{0}

\begin{table}
  \begin{center}
    \renewcommand{\arraystretch}{1.3}
    \begin{tabular}{| c | c | c | c | c | }
      \hline
      	& Pressure (mbar) & ${\rm Re}_r = r u_*/\nu$ & ${\rm Re}_\lambda = \lambda u_*/\nu$ & $\Theta = u_*^2/\left[ \left( \rho_s/\rho - 1 \right) gd \right]$ \\
      \hline
    		& $1020$	& $13.15$ (transitional)	& $1.40 \times 10^3$	& $0.081$ \\  
      this     & $500$	& $9.36$ (smooth)		& $5.50 \times 10^3$	& $0.079$ \\  
      study	& $100$	& $2.36$ (smooth)		& $1.75 \times 10^3$	& $0.025$ \\  
                 & $50$	& $1.64$ (smooth)		& $2.80 \times 10^3$	& $0.025$ \\
      \hline
      Mars	& $6$	& $0.58$ (smooth)		& $3.70 \times 10^3$	& $0.015$ \\
      \hline
    \end{tabular}
  \end{center}
\caption{
\textbf{Comparison of dimensionless numbers under experimental and Martian conditions.}
Parameter values corresponding to the experiments are provided in the manuscript and Table S2. For Mars, typical values were assumed: $\rho = 0.02$~kg/m$^3$, $\mu = 1.08 \times 10^{-5}$~Pa s, $d = 125$~$\mu$m, $u_* = 1$~m/s, $\rho_s = 2900$~kg/m$^3$, $g = 3.71$~m/s$^2$, and $\lambda = 2$~m. Bedform wavelength, $\lambda$, refers to that of the largest bedform type present under any given pressure condition (impact ripples at $1020$~mbar, drag ripples otherwise).
}
\end{table}
\begin{table}
  \begin{center}
    \renewcommand{\arraystretch}{1.3}
    \begin{tabular}{| c | c | c | c | c | }
      \hline
pressure $p$~(mbar)	& $1020$	& $500$	& $100$	& $50$ \\
      \hline
atmospheric 	& & & & \\
density $\rho_a$~(kg/m$^3$)		& $1.19$	& $0.61$	& $0.12$	& $0.06$ \\
      \hline
shear velocity  & & & & \\
$u_*$~(m/s)	& $0.41 \pm 0.08$	& $0.57 \pm 0.09$	& $0.72 \pm 0.04$	& $1.01 \pm 0.10$ \\
      \hline
aerodynamic & & & & \\
roughness length& $(1.978 \pm 0.975)$		& $(2.794 \pm 0.975)$		& $(4.613 \pm 0.975)$		& $(5.264 \pm 0.975)$ \\
$z_0$~(m)	& $\times 10^{-4}$ & $\times 10^{-4}$ & $\times 10^{-4}$ & $\times 10^{-4}$ \\
      \hline
threshold shear & & & & \\
velocity $u_{*t}$~(m/s)	& $0.278 \pm 0.002$		& $0.339 \pm 0.003$		& $0.667 \pm 0.005$		& $0.769 \pm 0.006$ \\
      \hline
impact-ripple & & & & \\
wavelength $\lambda_i$~(cm) & $5.0 \pm 0.2$	& $2.9 \pm 0.1$	& $4.1 \pm 0.2$	& $4.5 \pm 0.2$ \\
      \hline
impact-ripple & & & & \\
height (calculated)& $0.25$	& $0.15$	& $0.21$	& $0.23$ \\
$h_i$~(cm)	& & & & \\
      \hline
number of & & & & \\
measurements	& $80$		& $48$		& $42$		& $65$ \\
(small ripples)$^a$	& & & & \\
      \hline
drag-ripple & & & & \\
wavelength $\lambda_l$~(cm) & n/a		& $28.6 \pm 2.8$		& $36.5 \pm 3.6$		& $83.1 \pm 2.2$ \\
      \hline
drag-ripple	& & & & \\
height $h_l$~(cm)	& n/a		& $1.4 \pm 0.1$	& $2.1 \pm 0.2$	& $2.4 \pm 0.1$ \\
      \hline
number of 	& & & & \\
measurements	& n/a		& $6$	& $5$	& $5$ \\
(drag ripples)	& & & & \\
      \hline
modeled-normalized & & & & \\
effective roughness	& $0.60$	& $1.47$	& $1.20$	& $0.53$ \\
form drag$^b$ $z_0 u_*/\nu$	& & & & \\
      \hline
modeled-normalized	& & & & \\
effective roughness	& $3.94$	& $2.48$	& $0.90$	& $1.59$ \\
saltation$^c$ $z_0 u_*/\nu$	& & & & \\
      \hline
    \end{tabular}
  \end{center}
\caption{
\textbf{Summary of experimental parameters and derived measurements.}\\
$^a$Values represent the number of ripples measured from optical, high-resolution images.\\
$^b$Values represent the mean compound roughness from skin friction and form drag from all bedform scales present (Jia et al. \cite{jia2023} no saltation; Fig. 4).\\
$^c$Values represent the mean normalized aerodynamic roughness induced by saltation, modeled using the Jia et al. model \cite{jia2023} in an inverse fashion (Fig. 5).
}
\end{table}
%

%------------------------------------------------------------------
\newpage
\renewcommand\thefigure{S\arabic{figure}}
\setcounter{figure}{0}

\begin{figure*}
\centering
\includegraphics[width=0.85\linewidth]{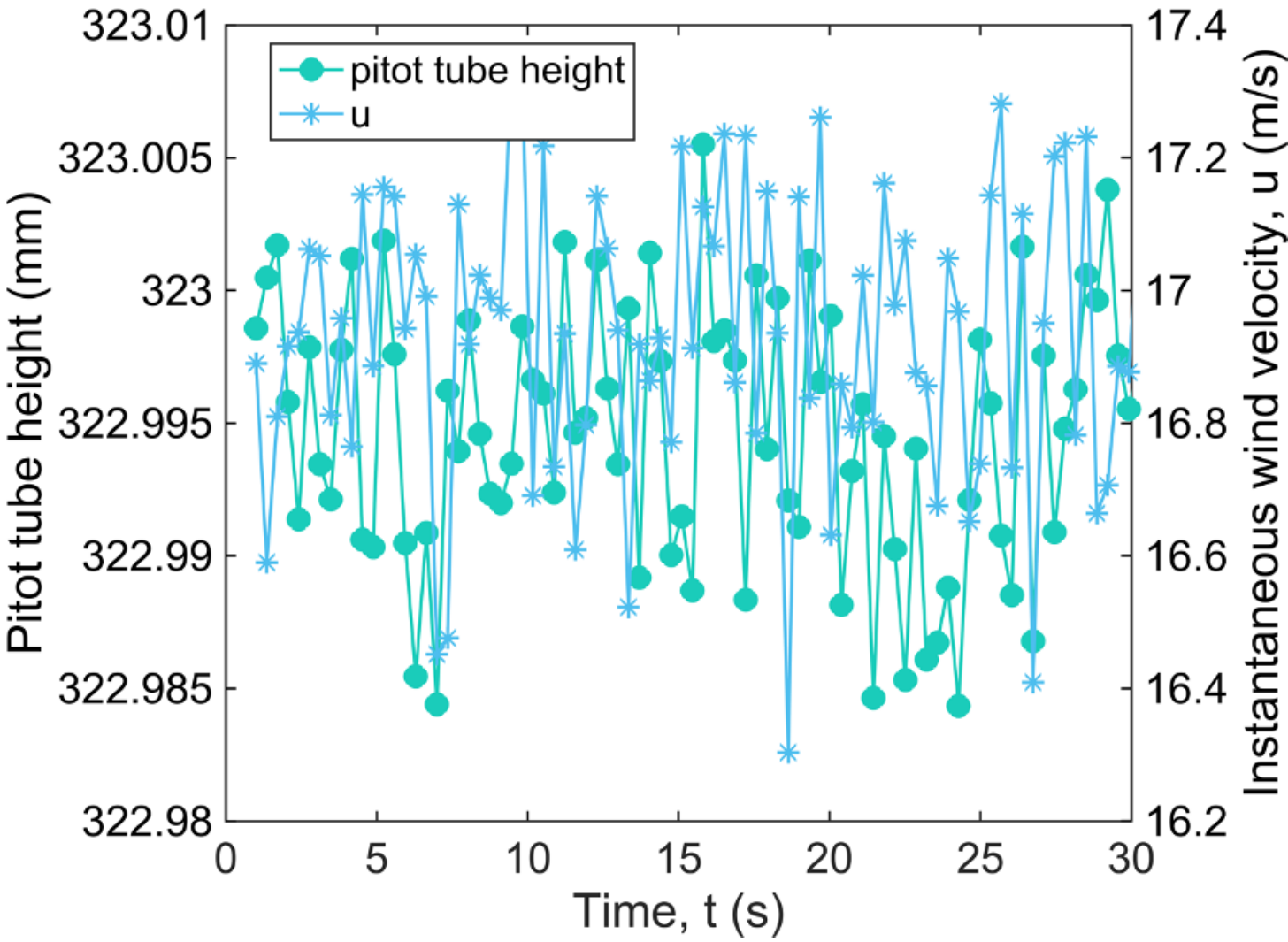}
\caption{
Example measured pitot tube height and wind speeds ($p = 50$~mbar). Measurements were made after ripples had equilibrated.
}
\end{figure*}

\begin{figure*}
\centering
\includegraphics[width=0.5\linewidth]{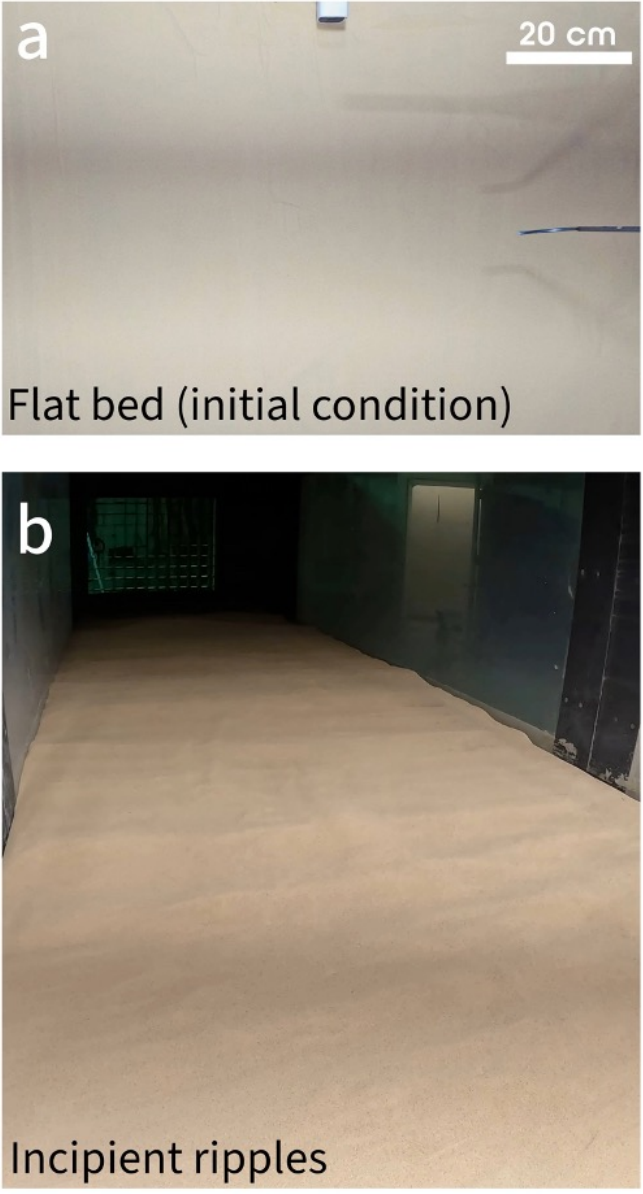}
\caption{
Sediment bed in the MARSWIT test section. (a) Top-down view of the test section showing the initial condition of the bed. The picture shows a portion of the Pitot tube (right) and a high-resolution infrared camera (top) used for side-view control of the bed evolution. (b) Incipient bedforms forming at $500$~mbar after $\simeq 10$~min (winds directed toward the camera). The wind tunnel is $1.3$~m wide.
}
\end{figure*}

\begin{figure*}
\centering
\includegraphics[width=0.75\linewidth]{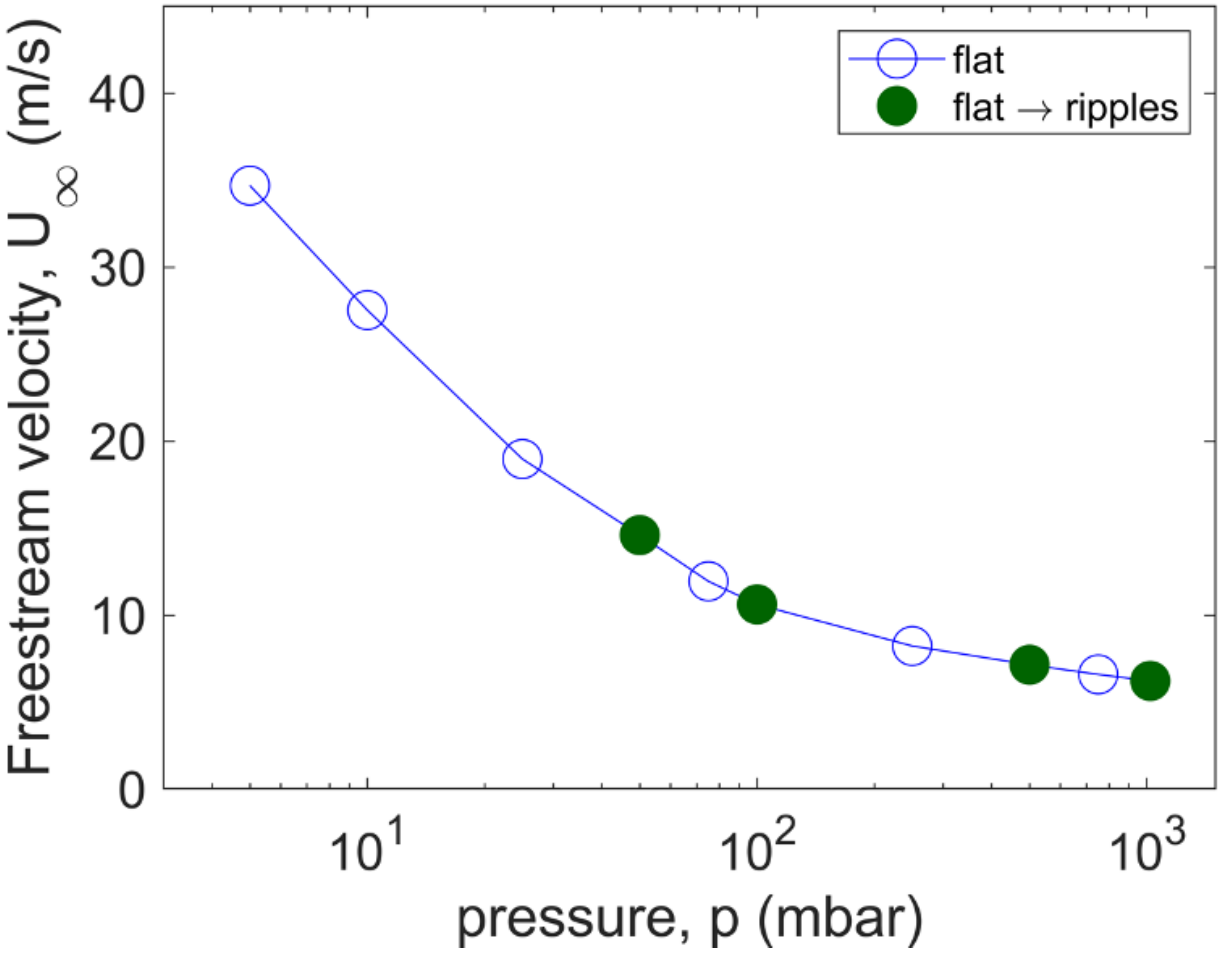}
\caption{
Threshold freestream velocity, $U_\infty$, for continuous saltation as a function of pressure. Open symbols indicate measurements performed over a flat bed; filled symbol reflect measurements over equilibrated rippled beds. 
}
\end{figure*}

\begin{figure*}
\centering
\includegraphics[width=0.75\linewidth]{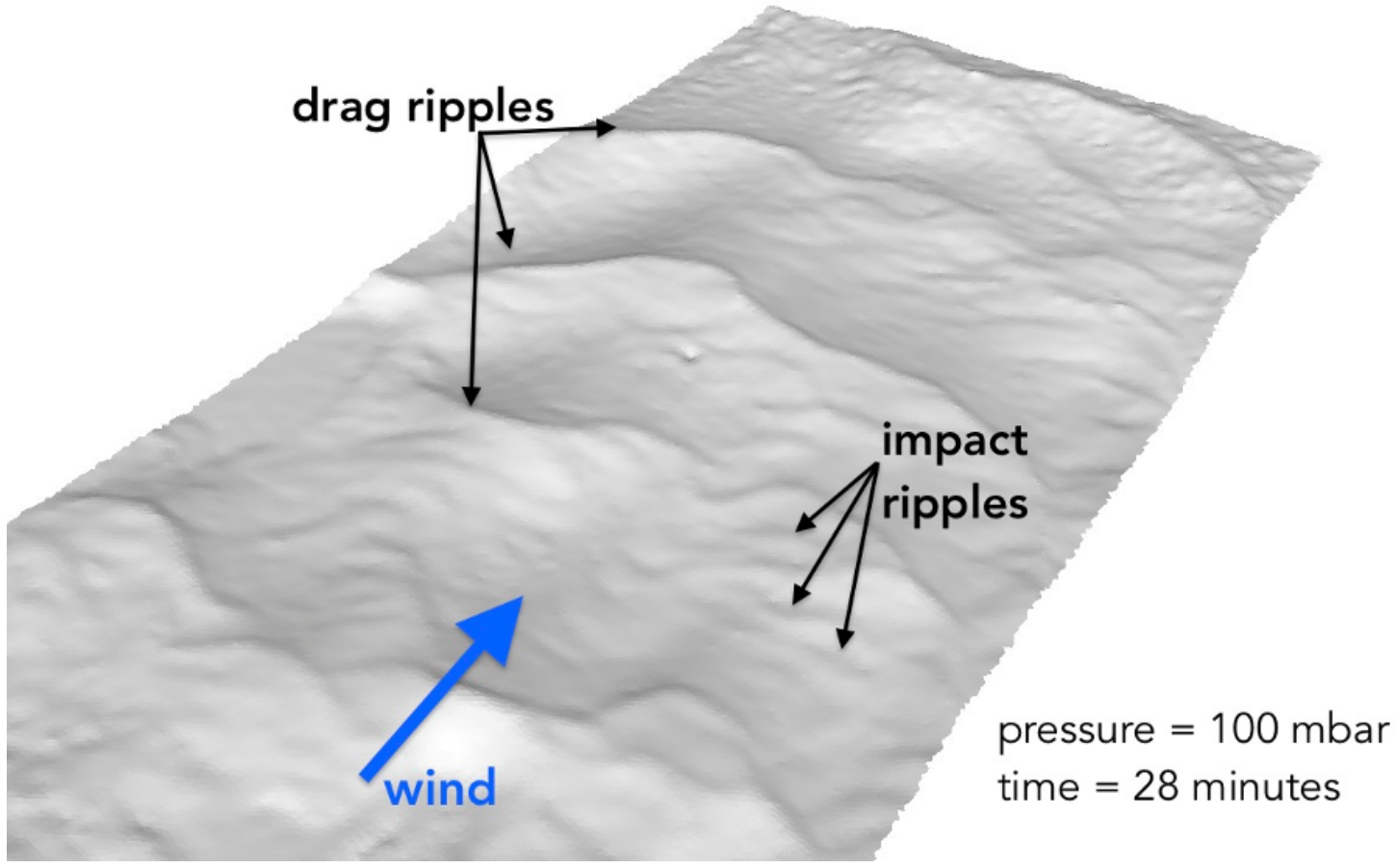}
\caption{
Example 3D scan of equilibrated large and impact ripples after $28$~min of evolution from a flat bed at $100$~mbar. The scan is approximately $0.92$~m across.
}
\end{figure*}

\begin{figure*}
\centering
\includegraphics[width=0.75\linewidth]{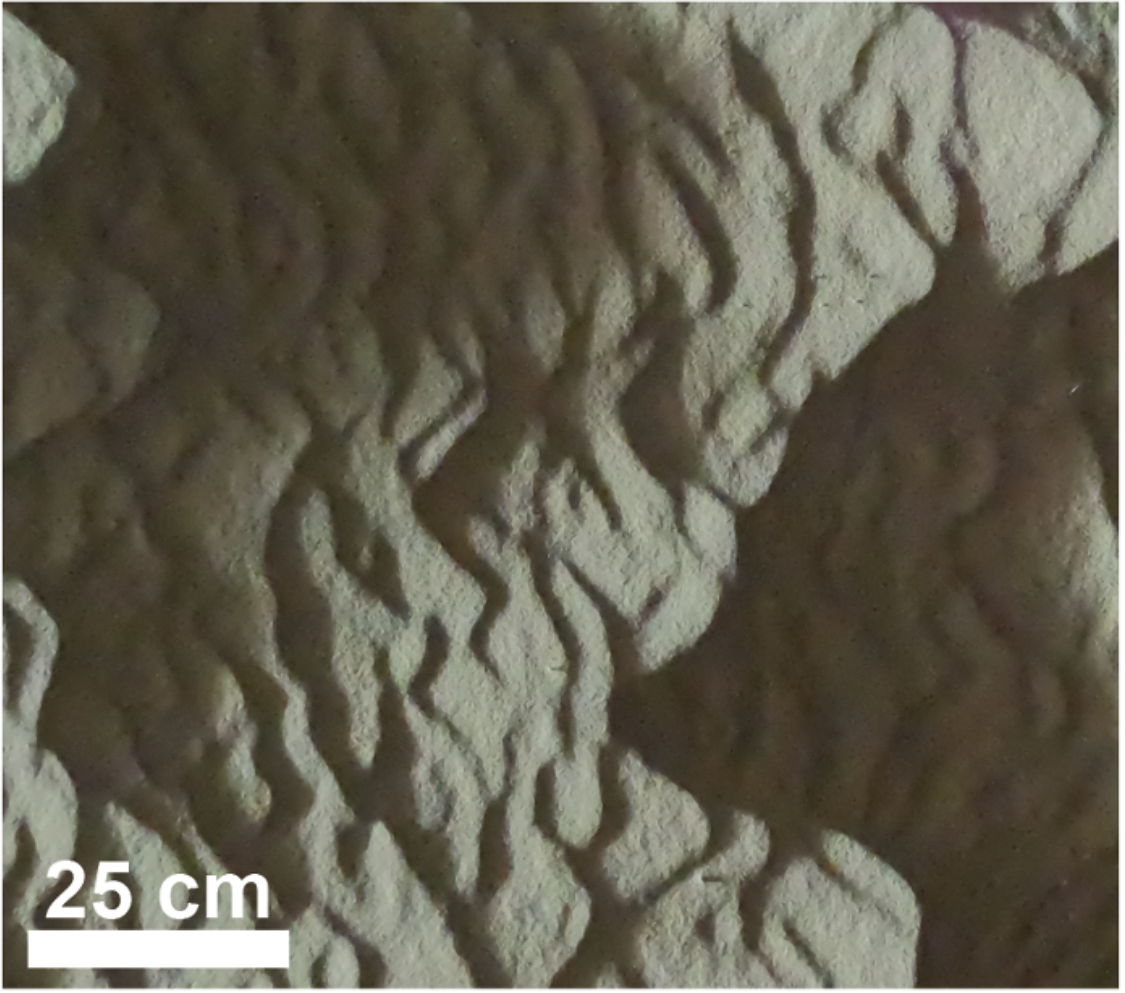}
\caption{
Example image used to reconstruct impact-ripple height from shadow lengths, $l$, at $50$~mbar. Ripple height was calculated as $h_i = l \tan \left( \alpha_i+ \alpha_s \right)$, with $\alpha_i$ the incidence angle of light and $\alpha_s$ the stoss slope angle of large ripples.
}
\end{figure*}

\begin{figure*}
\centering
\includegraphics[width=0.85\linewidth]{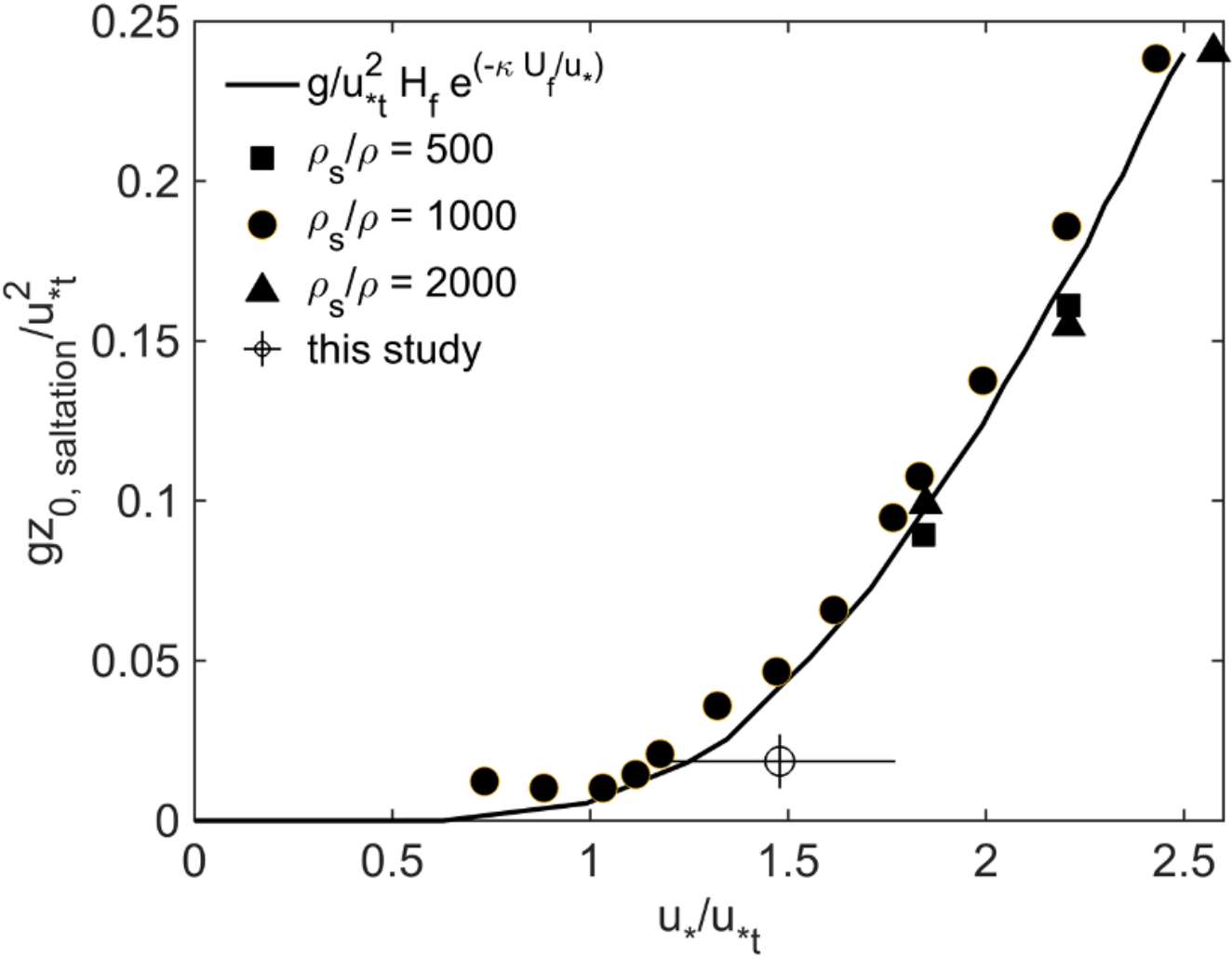}
\caption{
Dimensionless aerodynamic roughness due to saltation, $z_{0, \, {\rm saltation}}$, as a function of $u_*/u_{*t}$. Solid symbols denote predictions from the numerical model of Dur\'an et al. \cite{duran2011}, and the solid line represents $z_{0, \, {\rm saltation}} = H_{\rm f} \exp \left( -\kappa U_{\rm f}/u_* \right)$, where focal height and velocity were calculated using an analytical approximation \cite{duran2011}.
}
\end{figure*}

\begin{figure*}
\centering
\includegraphics[width=0.85\linewidth]{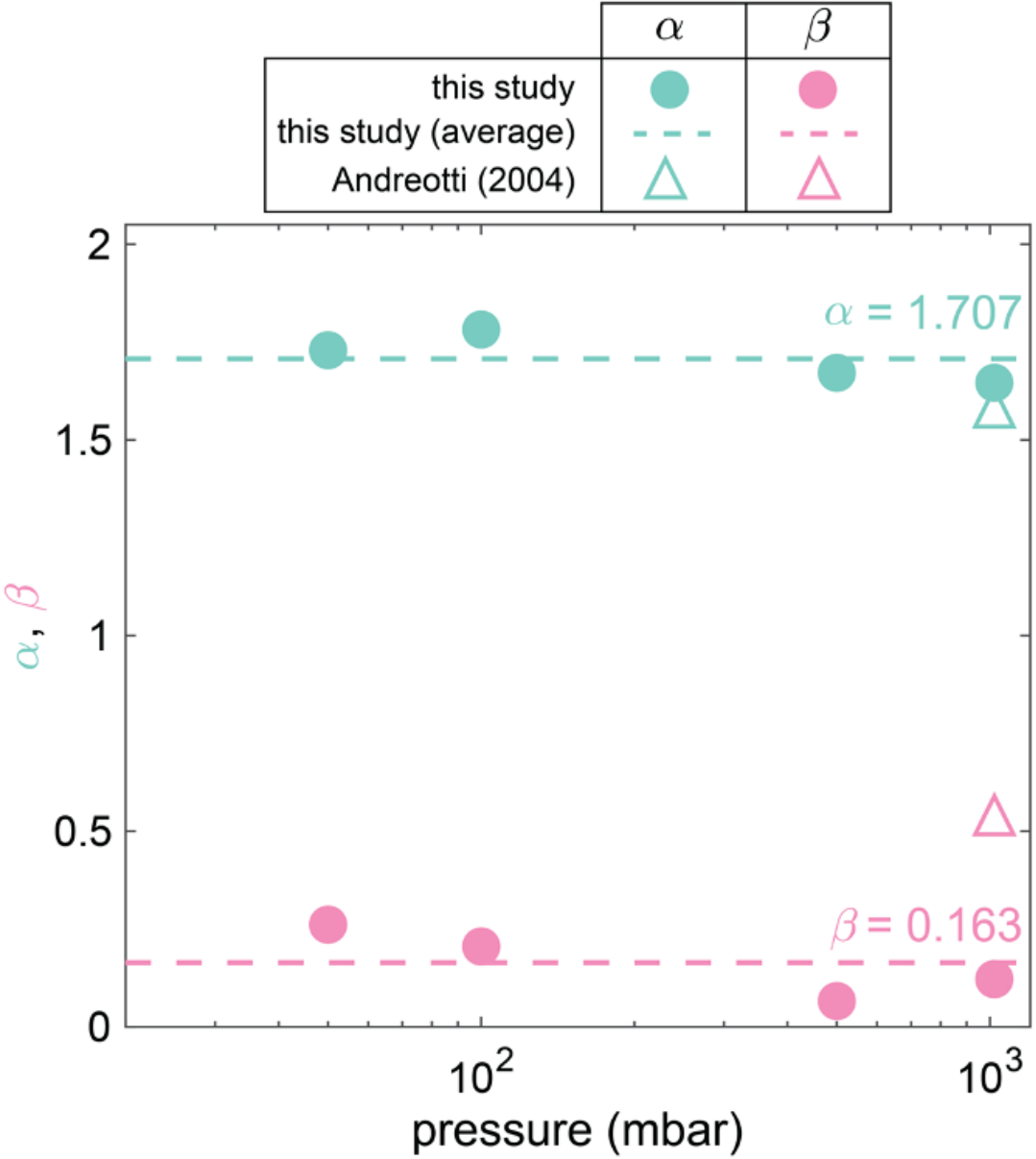}
\caption{
Best-fit coefficients, $\alpha$ and $\beta$, at different pressures and averaged values. Coefficients of Andreotti (2004) at $1020$~mbar are added for comparison.
}
\end{figure*}

\begin{figure*}
\centering
\includegraphics[width=0.85\linewidth]{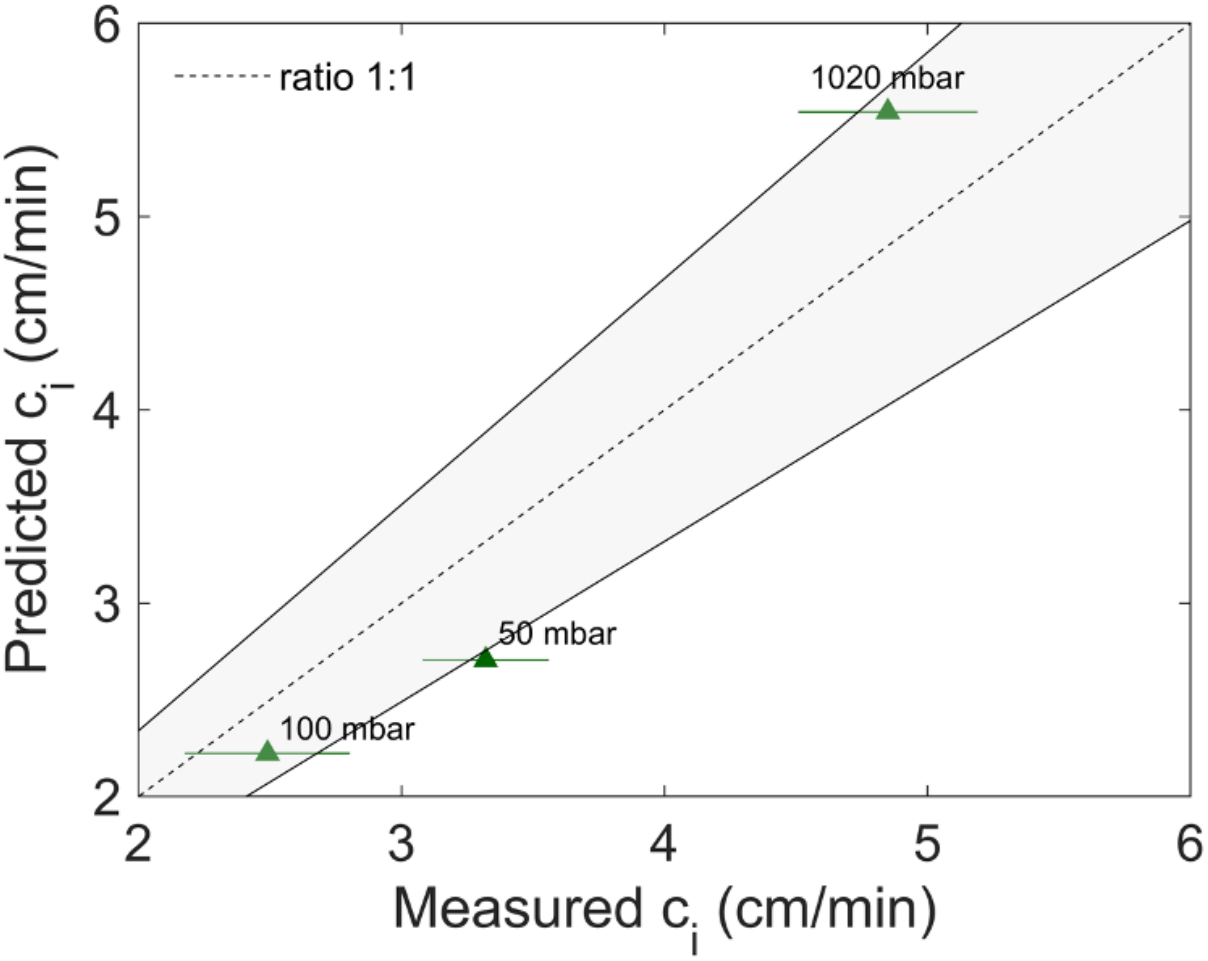}
\caption{
Predicted vs. measured migration rates, $c_i$ of small, impact ripples. Measured values were derived from postprocessing of top-down movies recorded during experiments. Shaded area indicates an error of $\pm 17$\%. Predicted values were calculated as $c_i=0.1 \sqrt{(\rho_a/\rho_s \left[ (u_*/u_{*t} )^2-1 \right]} ) u_{*t}$. Impact ripples forming at $500$~mbar were not clearly visible in the IR footage, such that their migration rate could not be measured.
}
\end{figure*}

\clearpage
\newpage
%________________________________________________________________________
% Bibliography
\bibliographystyle{elsarticle-num}
\bibliography{marswit}

\end{document}